\documentstyle[psfig]{l-aa}

\begin{document}

\thesaurus{ 02 (02.04.1; 02.05.2; 02.16.1), 08 (08.23.1)}

\title{Three-body crystallization diagrams and the cooling of white dwarfs}

\author{Laurent SEGRETAIN}

\institute{Centre de Recherche Astronomique de Lyon
(UMR 142 du CNRS), Ecole Normale Sup\'erieure de Lyon, 46 all\'ee d'Italie,
F-69364 Lyon C\'edex 07, France}

\date{ Received August 11; accepted October 16, 1995}

\maketitle

\begin{abstract}
The 3-body crystallization diagrams of C/O/Ne ionic mixtures
characteristic of white dwarf interiors are
examined within the framework of the density-functional theory of
freezing. The crystallization process is described more accurately than
in former calculations where the three-component system was treated as
an effective two-component mixture (Segretain et al. 1994). The
distillation process due to neon-crystallization is found to occur only
for the late stages of crystallization. At the beginning, the presence
of neon plays only a minor role and the phase diagram resembles a pure
carbon-oxygen diagram.  The final phase diagram is found to exhibit an
azeotropic point with a neon concentration $x_{Ne}=0.22$, a carbon
concentration $x_{C}=0.78$ and an oxygen concentration $x_O=0$, so that
during the distillation process, the fluid crystallizes into a pure
neon-carbon solid.  The critical temperature is $T_A=0.85\,T_C$, where
$T_C$ is the pure carbon crystallization temperature. We use this
accurate phase diagram to calculate the total gravitational energy
released during white dwarf crystallization and the related time delay.
The final result yields $\Delta \tau \approx 2.6$ Gyr, among which
about 20\% are due to the neon-distillation process.


\end{abstract}

\section{Introduction}

Following the pionneer work of Stevenson (1980), Garcia-Berro et al.
(\cite{GarcHernMochIser88}) have demonstrated the importance of the
shape of the crystallization diagram on the cooling history of white
dwarfs (WD). In the eighties, all the work has focussed on the
crystallization of a pure carbon-oxygen mixture (Stevenson
\cite{Stev80}; Mochkovitch \cite{Moch83}; Barrat et al.
\cite{BarrHansMoch88}; Ichimaru et al.  \cite{IchiIyetOgat88}). More
recently, Isern et al.  (\cite{IserMochGarcHern91}) have pointed out
the importance of minor elements, in particular neon, on the general
cooling process.  In order to examine in details the effect of these
different crystallization processes, Segretain \& Chabrier
(\cite{SegrChab93}) have calculated the crystallization diagrams of
dense {\it binary} ionic mixtures with arbitrary charge ratios under
thermodynamic conditions characteristic of dense stellar plasmas. The
diagram was found to evolve from a spindle form to an azeotropic, and
eventually an eutectic form, with increasing charge ratio $Z_2/Z_1$.
These binary diagrams have been used by Segretain et al.
(\cite{SegrChabHernGarcIserMoch94}) to compute the cooling of a
C/O/Ne/Fe WD. The crystallization processes were shown to bear
major consequences on the cooling time and the luminosity function of
old WDs (Segretain et al. 1994; Hernanz et al. 1994).
 In these calculations, and given the complexity of the calculation of
multi-component phase diagrams, the ternary C/O/Ne or C/O/Fe mixtures
were assumed to behave as {\it effective binary} mixtures composed of
neon (resp. iron) and an element of average charge $\langle
Z\rangle=7$.  The calculations showed that neon plays a major role in
the cooling of WD, because of a distillation process bound to the
presence of an azeotropic point in the phase diagram. The resulting
sedimentation of neon was found to have an effect on the cooling time
comparable to the one due to the crystallization of the carbon-oxygen
mixture itself. In this paper we have extended the calculations of
Segretain \& Chabrier (\cite{SegrChab93}) to the calculation of a true
{\it ternary} C/O/Ne diagram. The diagram is calculated within the
framework of modern density functional theory (DFT) of freezing, and
the plasma correlation functions are calculated
 within the mean spherical approximation (MSA) (see Segretain \&
 Chabrier 1993 for details). In Sect. 3, we describe the resulting
diagram and we compare it with the previous effective two-body diagram.
In Sect. 4, we use this ternary phase diagram to derive the global
effect of neon crystallization on the cooling of a WD.

\section{Method}

The crystallization diagram has been computed within the framework of
the density functional theory (DFT) of freezing.  One of the main
advantages of the DFT is that we can calculate directly the free energy
{\it difference} between the solid and the fluid phases. We consider a
mixture of $N_i$ ions of charge $Z_i$ ($i=1,2,3$)  in a volume V at a
temperature T. The number densities and molar fractions read
respectively $\rho_i=N_i/V$ and $x_i=N_i/N$ (i=1,2,3) with $N=\sum_i
N_i$.  The electronic and ionic coupling parameters are defined
respectively as $\Gamma_e=e^2/a_ek_BT$ and
$\Gamma_i=Z_i^{5/3}\Gamma_e$, where $a_e=(4\pi/3 N_e/V)^{-1/3}$ is the
mean inter-electronic distance and $<Z>=\sum_i x_iZ_i$ is the average
charge. The free energy of the liquid phase reads:

\begin{equation}
F_L=\sum_{i=1}^3 x_i \left( \ln x_i {Z_i\over <Z>} +
F^{OCP}(\Gamma_i) \right)
\end{equation}

where the first term on the r.h.s. is the ideal ionic contribution
whereas the second term denotes the correlation terms. This term is
well approximated by a linear interpolation of the free energies of the
pure phases (designated by the superscript OCP -One Component Plasma-),
for which analytical fits are available (Stringfellow et al.
\cite{StriDeWiSlat90}).  The free energy of the solid writes:

\begin{displaymath}
F_S=F_L+\Delta F
\end{displaymath}
\begin{displaymath}
\Delta F=\Delta F^{id}+\Delta F^{ex}
\end{displaymath}

with
\begin{equation}
{\Delta F^{id}\over NkT} = \sum_{\nu=1}^3 {\rho_{\nu_S} \over \rho_L}
\left\{ \ln \left[{(\alpha_\nu/\pi)^{3/2} \over \rho_{\nu_L} }
{\rho_{\nu_S} \over \rho_S} \right] - {3 \over 2} \right\}
\end{equation}
and
\begin{eqnarray}
\lefteqn{{\Delta F^{ex}\over NkT} = -{1\over 2}\sum_{\vec{G}}'
\sum_{\nu,\mu=1}^3 {\rho_{\nu_S} \rho_{\mu_L} \over \rho_L^2} C_{\nu
\mu}(\vec{G})\times} \hspace{40mm}\nonumber\\
& & \exp\left[-{G^2 \over 4} \left({1\over \alpha_\nu} + {1\over
\alpha_\mu}\right)\right]
\end{eqnarray}

where $\sum_{\vec{G}}'$ represents the sum over all the reciproqual
lattice vectors (RLV) for $\vec{G}\not=0$. The $\rho_{\nu}$ denote the
number density of the $\nu$ component either in the solid (S) or in the
liquid phase (L).  For each concentration $x_\nu$ at a given
temperature T, $\Delta F$ is minimized with respect to the width of the
gaussians $\alpha_\nu$ in order to calculate $F_S$. The $C_{\mu
\nu}(\vec{G})$ are the Fourier transforms of the direct correlation
functions of the liquid near crystallization. These latter are computed
within the framework  of the Mean Spherical Approximation (MSA).  As in
Segretain \& Chabrier (\cite{SegrChab93}), we use a reduction of the
second RLV in $C_{\mu \nu}(\vec{G})$ in order to obtain a bcc crystal
for the one component system at $\Gamma_e Z^{5/3}=178$ (Slattery et
al.  \cite{SlatDoolDeWi82}; Stringfellow et al.
\cite{StriDeWiSlat90}).

\section{C/O/Ne phase diagram}

Since the pressure of the degenerate electron gas is considerably
larger than the ionic pressure under the conditions of interest
(degenerate stars interiors), fluid-solid phase equilibrium at constant
pressure and temperature is equivalent to phase coexistence at constant
electron density $\rho_e$ and temperature T, i.e. at constant
$\Gamma_e$. For fixed values of $\Gamma_e$ and T, the concentrations of
coexistent fluid and disordered solid phases are determined by usual
double-tangent constructions.

\begin{figure}
\psfig{file=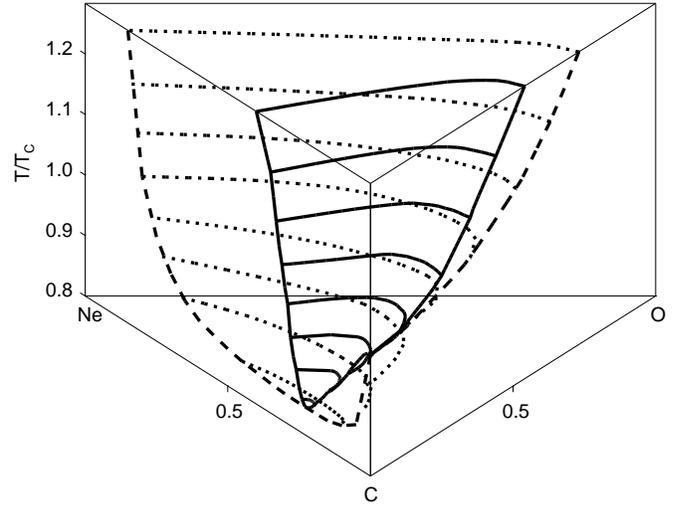,width=\linewidth}
\caption[]{C/O/Ne crystallization diagram for $\Gamma_e\geq 7$. The
coordinates of the azeotropic point are $T_a=0.85\,T_C$ and $x_C=0.78$,
$x_O=0.0$, $x_{Ne}=0.22$, where $T_C$ is the crystallization
temperature of pure carbon. The bold solid lines visualize the {\it
liquidus} and the bold dotted and dashed lines the {\it solidus}.
}
\end{figure}

Figure 1 shows the resulting 3-component phase diagram for the C/O/Ne
mixture under thermodynamic conditions characteristic of WD interiors,
i.e.  $10^{10}\leq P \leq 10^{16}$ Mbar,
$10^3\leq\rho\leq10^6\,\hbox{gcm}^{-3}$ and for
$T\leq5.10^7\,\hbox{K}$, which yields $\Gamma_e\geq 7$.  The 3-D
picture of the C/O/Ne diagram can be first visualized with the 2-D
projections on the C/O, C/Ne and O/Ne planes. Note that, for sake of
clarity, only the lower part of the general diagram ($T/T_C\le 1.2$,
where $T_C$ is the crystallization temperature of pure carbon), is
shown. The pure carbon and oxygen mixture ($X_{Ne}=0$) limit
crystallize into a spindle diagram, as expected from the small charge
ratio. So does the bare O/Ne mixture ($X_{C}=0$ limit), with a larger
crystallization temperature, since $T_{Ne}>T_{C}$ and $T_O>T_C$ where
$T_i\,,\, i=C,O,Ne$, denotes the crystallization temperature of pure
carbon, oxygen and neon respectively. The pure C/Ne ($X_{O}=0$) diagram
exhibits an azeotropic behavior with an azeotropic point for
$x_C=0.78$, $x_{Ne}=0.22$ and $T_a=0.85\,T_C$. The true 3-body diagram
is visualized by the bold lines, where the solid line display the {\it
liquidus} and the dotted and dashed lines represent the {\it solidus}.
As shown, the azeotropic point of the {\it three}-body diagram is the
azeotropic point of the {\it two}-body C/Ne mixture ($X_O=0$), which is
the lowest stable point for the liquid in the three-component C/O/Ne
mixture.

We also note that, for small concentrations of neon, and for
temperatures well above the azeotropic temperature, the liquid and the
solid have the same Ne-concentration. The mixture then crystallizes like
an equivalent Ne-enriched C/O mixture. In other words, at high
temperature, and for low concentrations of neon, the crystallization
diagram is not affected by the presence of neon. As the temperature
approaches the azeotropic temperature, neon starts to modify the
crystallization process. The distillation process, as described in
Segretain et al. (\cite{SegrChabHernGarcIserMoch94}), begins. The
liquid becomes enriched in neon and the diagram evolves into the
azeotropic shape.

This behavior is different from the one obtained with the {\it
effective} 2-body mixture used in Segretain et al.
(\cite{SegrChabHernGarcIserMoch94}), where the Ne-enriched process was
found to start as soon as crystallization set in. As described above,
the present complete 3-body calculations show that C/O crystallization
first takes places without Ne-enrichment.

This bears important consequences for the astrophysical application, as
will be shown in the next section, where the diagram will be examined
for C/O/Ne concentrations characteristic of WD interiors.

\section{Consequences on white dwarf cooling}

The mass fractions characteristic of WD interior are $X_{Ne}=1.0$\% and
$X_{C}=X_{O}=49.5$\%, uniformly distributed in the liquid phase (the
effect of an initial non-uniform profile has been examined in details
in Segretain at al. 1994). The exact C/O mass fraction in WD interiors is
subject to important
uncertainties. The present equimolar distribution has been chosen for the sake
of simplicity.
Larger initial oxygen concentrations in the core of the WD will, at any
rate, lead to even smaller energy release by Neon sedimentation.
As shown on the diagram of Fig. 1, the
process of crystallization develops as follows. Crystallization begins
at the center of the star where the density, and thus the
crystallization temperature, is the largest
($T_C\propto\rho^{1/3}/178$).  In that case, as mentioned in the
previous section, we get the same behaviour as for a pure C/O mixture,
but with concentrations $X_C=X_O=0.495$, i.e the same neon
concentration in the solid and the liquid phase.  Until $T/T_C=1.12$
(point A in Fig. 2) the resulting sedimentation of oxygen is the same
as in a pure C/O WD. Beyond $T/T_C=1.12$, the distillation
process begins, so that neon concentration becomes different in the two
phases.  However, the neon effect on the release of gravitational
energy will be small because, at point A, about $70$\% (in mass) of the
star has already crystallized. This is very different from the result
obtained with the {\it effective} 2-body diagram, where neon was found
to crystallize first. At the end of the distillation process, i.e. at
the azeotropic temperature (point B in Fig. 2), all neon has
crystallized; the remaining liquid contains only carbon and oxygen and
sedimentation of oxygen continues normally. The oxygen and neon
profiles in solid along the crystallization process are shown on Fig
2.

\begin{figure}
\psfig{file=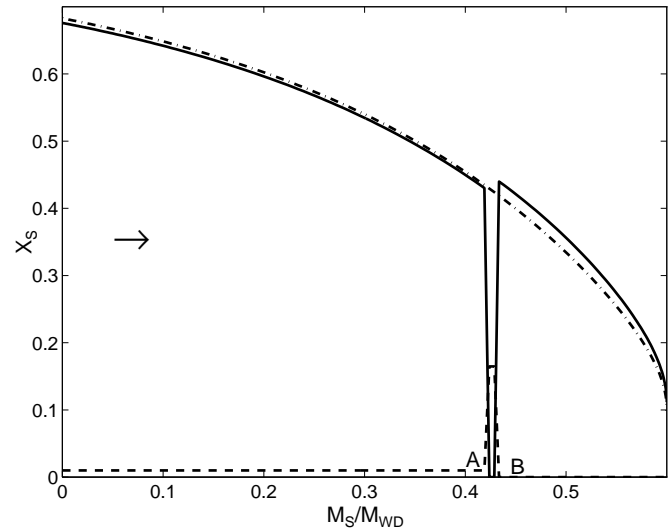,width=\linewidth}
\caption[]{Concentration profiles at the end of the crystallization,
for two different WDs. Dot-dashed line : oxygen profile for a pure C/O
WD. Solid line : oxygen profile for a C/O/Ne WD.  Dashed line : neon
profile for a C/O/Ne WD.  The crystallization front evolves as
indicated by the arrow.  The neon-distillation process begins at point
A and ends at point B.  During this process, the solid which forms does
not contain oxygen, since the azeotropic point is characterized by
$X_O=0$. After point B, all neon has crystallized; the remaining solid
contains only carbon and oxygen, and the usual oxygen sedimentation
process proceeds again.
}
\end{figure}

These concentration profiles are used to compute the evolution of the
binding energy along crystallization, as in Segretain et al. (1994).
The difference of binding energy due to crystallization is
 $\Delta B=3.10\,10^{46}$ erg, only $\sim$20\% more than the one
obtained for the crystallization of a pure C/O WD. The effect
of neon crystallization, when properly taken into account, is then
smaller than when treated within the effective 2-body approximation.
The release of binding energy due to the crystallization of the C/O/Ne
mixture yields a total time delay about 20\% larger than the one due to
the crystallization of a pure C/O mixture, i.e $\Delta \tau \approx
2.6$ Gyr. The effect intrinsically due to neon crystallization is about
0.4 Gyr. This effect on the cooling time, however, does not modify the
age of the Galactic disk, based on the calculation of the WD
luminosity function, obtained by Hernanz et al.
(\cite{HernGarcIserMochSegrChab94}), as explain in section 3.2 of the paper of
Hernanz et al. (\cite{HernGarcIserMochSegrChab94}).

\section{Conclusion}

In this paper we have computed for the first time the crystallization
diagram of a {\it three-component} C/O/Ne mixture characteristic of WD
interiors. The general phase diagram is found to be of the azeotropic
form. We use this phase diagram to compute the crystallization process of a
uniform distibution C/O/Ne WD with $X_{Ne}=1.0$\% and
$X_{C}=X_{O}=49.5$\%.
The presence of neon is found to be inconsequential at the
beginning of the crystallization process. The neon concentration
remains the same in the solid and in the liquid phase, so that the
major process is the sedimentation of oxygen, characterized by a
spindle diagram, as found for a pure C/O WD.  When 70\% (in mass) of
the WD has crystallized, the shape of the diagram evolves
substantially, and the influence of neon becomes important. A
distillation process begins and all the neon left in the liquid
crystallizes at the azeotropic concentration/temperature, i.e
$T_a=0.85\,T_C$, $x_C=0.78$, $x_O=0.0$ and $x_{Ne}=0.22$.  This
crystallization process is different from the one obtained with the
{\it effective} 2-body diagram for which the distillation process is
found to begin as soon as crystallization sets in.  The time delay
produced by the afore-mentioned general crystallization process is
about 2.6 Gyr, i.e about 20\% larger than the one obtained for a pure
C/O mixture. The effect of neon is only of 0.4 Gyr.  These detailed
calculations show that, though not completely negligible, the effect of
neon, and iron as well, does not bear important
consequences on the general cooling history of WDs.

\begin{acknowledgements}
The author are grateful to G. Chabrier for helpful discussions.
\end{acknowledgements}


\end{document}